\documentclass[aps,prl,reprint,amsmath,amssymb,amsfonts,groupedaddress]{revtex4-2}

\usepackage{graphicx}
\usepackage{dcolumn}
\usepackage{bm}

\begin{document}

\preprint{APS}

\title{Unified Perspective on Single Cyclotron Electron with Radiation-Reaction from Classical to Quantum}


\author{Qiang Chen}
\email{corresponding author: qiangchen@zzu.edu.cn}
\affiliation{National Supercomputing Center in Zhengzhou, Zhengzhou University, Zhengzhou, Henan 450001, China}

\author{Peifeng Fan}
\email{corresponding author: pffan@szu.edu.cn}
\affiliation{Key Laboratory of Optoelectronic Devices and Systems, College of Physics and Optoelectronic Engineering, Shenzhen University, Shenzhen, Guangdong 518060, China}
\affiliation{Advanced Energy Research Center, Shenzhen University, Shenzhen, Guangdong 518060, China}

\author{Jianyuan Xiao}
\email{corresponding author: xiaojy@ustc.edu.cn}
\affiliation{School of Nuclear Science and Technology, University of Science and Technology of China, Hefei, Anhui 230026, China}


\date{\today}

\begin{abstract}
We show a unified physical picture of single cyclotron electron with radiation-reaction, which bridges the 
classical electron models and quantum mechanical self-consistent field theory. On a classical level, we suggest 
an improved electrodynamical action, which build the classical electron models into a first-principle framework. 
The link between dynamical defections and non-physical action configurations emerges naturally. On a quantum level, 
a self-consistent description for electron gyro-motion with self-force is constructed in the Schr\"odinger-Maxwell 
theory. We derive a class of asymptotic equations. The leading and next-to-leading orders give a good analogue of a 
classical cyclotron electron, and the limit field theory avoids classical electron induced defections gracefully. 
Beyond the Hamiltonian perturbation theory, we use state-of-the-art geometric simulator to observe single 
electron gyro-motions at quantum region. The non-linear and non-perturbative features captured by simulations 
provide a complete physical picture in a very wide range. We show an optimal complementary relation between 
classical and quantum cyclotron electrons, and find a strange and inexplicable electron chimera state existing 
at strong non-linear regions, which may be observed in astrophysical environments and strong magnetic experiments.
\end{abstract}


\maketitle

\paragraph{Introduction}---
\label{pa:1}
Cyclotron electrons in a strong magnetic field play a central role in many branches of modern physics 
and associated advanced technologies\cite{Jackson1962,Lai2001,Chu2004}. Let us look into the deep universe. 
On the surface of an X-ray pulsar, the typical magnetic fields are on the order $B\sim10^{12}$ G. Immersed 
in such an extreme magnetic field, the electrons in magnetosphere plasmas exhibit strong anharmonic cyclotron 
absorption features observed in astrophysical spectra data\cite{Michel1982,Piran2005}. Although some of these 
features have been explained as inelastic electron-photon scatterings and relativistic effects, more mechanisms 
hidden in the cyclotron absorption lines remain to be studied\cite{Chen2021}. To achieve a big dream that one day 
humans have access to interstellar journey, physicists and engineers started a long march to controlled fusion for 
more than half a century. In a magnetic-confinement fusion reactor, e.g. a tokamark or stellarator, the fusion plasmas 
are well bounded on a compact $\mathbb{T}^{2}$ manifold, where the helical motions of confinement electrons affect 
their equilibrium, stability, transportation and relaxation, which finally determine the quality of fusion plasmas\cite{Boozer2005}. 
Both single electron dynamics and gyrokinetics are introduced to explain these complex electron gyro-motions, but 
more advanced theoretical tools are needed\cite{Boozer2005,Brizard2007}. Recent big fusion experiments are 
strongly supported by modern vacuum electronic technologies. The core heaters used in \emph{ITER} project 
are MW grade gyrotrons covering 110, 140 and 170 GHz frequencies. In the tube, an intense electron beam is 
modulated by a T grade magnetic field, which transfer energy from cyclotron electrons to high power microwaves, 
where the cyclotron radiation and radiation damping are most important problems\cite{Kikuchi2012}. In above three 
fields, researchers treat cyclotron electrons with different theories and approaches. Sometimes, they are thinked 
as point-like objects in maths which are governed by the variational principle\cite{Jackson1962,Landau1975}. Sometimes, 
they are described as a complete fluid and associated waves emerge\cite{Boozer2005}. Sometimes, they are given different 
shapes and finite volumes to avoid inexplicable divergency and bad causality which even exist in the quantum electrodynamics 
(QED) category\cite{Levine1977,Blanco1986,Harte2006,Moniz1974,Moniz1977,Higuchi2002,Higuchi2004,Higuchi2009,Dinu2016}. 
Sometimes, they are picked up from the scattering amplitudes as elements in an abstract Hilbert space\cite{Feynman1962}. 
Though the images of an electron appear in these fields are very different, they can successfully describe the properties 
of an electron in relevant phenomena. As such, despite the QED has achieved great success in fundamental photon-electron 
interactions, an interesting question can be asked: \emph{What is a classical cyclotron electron}? To answer this question, 
we establish a unified physical picture for a cyclotron electron with radiation-reaction (R-R): \emph{Dressed magnetic 
coherent state bridges classical electron models and quantum mechanical self-consistent field theory}. We give a detailed 
discussion on the link between effective theories of the classical R-R and asymptotic theory of the Schr\"odinger-Maxwell 
(S-M) self-consistent field. With the help of an advanced numerical tool, we obtain an optimal complementary relation between 
classical and quantum cyclotron electrons, and find a new quasi-steady electron state existing at strong non-linear regions, 
which is recognized as a coherent-chaotic chimera state.

Our physical picture of a cyclotron electron can be used to unify different models and perspectives appear in plasma 
physics, astrophysics, accelerator physics and vacuum electronics. The strange and inexplicable electron chimera state 
in a strong magnetic field may be observed in experiments.

\paragraph{Classical electron model}---
\label{pa:2}
In classical electrodynamics, a cyclotron electron gets a self-force because of R-R, which introduces some fundamental 
difficulties, such as self-energy divergence, runaway and preacceleration\cite{Jackson1962,Landau1975}. These difficulties 
root in defective classical electron models, which treat a electron as a charged point-like object or small rigid body. 
Lorentz and Abraham (A-L) first gave an estimate on the R-R force via the averaged Larmor power and derived the famous 
A-L equation\cite{Lorentz1892,Abraham1905},
\begin{eqnarray}
\dot{\bm{p}} = \bm{F}_{ext} + \bm{F}_{R},\label{eq:al}
\end{eqnarray}
where the R-R force $\bm{F}_{R}=\tau_{0}m_{e}\dddot{\bm{x}}$ and transition time $\tau_{0}=\frac{2e^{2}}{3m_{e}c^{3}}$. 
All non-physical defections inherent in the A-L formula come from the 3rd-order jerk term which breaks the causality and 
time-reversal symmetry. Thus future signals of the external force $\bm{F}_{ext}$ affect the current electron acceleration 
constantly. Then runaway and preacceleration occur. Though pseudophysics exist, the A-L theory is regarded as a precise model 
on which many researchers construct their theories based. At relativistic region, Dirac extended it into a covariant form\cite{Dirac1938},
\begin{eqnarray}
\frac{{\rm{d}}P^{\mu}}{{\rm{d}}\tau}=\frac{e}{m_e}F^{\mu\nu}P_{\nu}+\tau_{0}\left[\frac{{\rm{d}}^{2}P^{\mu}}{{\rm{d}}\tau^{2}}-\frac{P^{\mu}P_{\nu}}{m^2_ec^{2}}\frac{{\rm{d}}^{2}P^{\nu}}{{\rm{d}}\tau^{2}}\right],\label{eq:ald}
\end{eqnarray}
where $F^{\mu\nu}$ is the electromagnetic 2-form, $P^{\mu}$ and $\tau$ are the 4-momentum and proper time respectively. There 
is even a hybrid QED extension constructed by stochastic dynamics and field theory, which is named as Abraham-Lorentz-Dirac-Langevin 
equation\cite{Johnson2002}. We emphasize that the dynamical defections can not be removed entirely in all these models. To overcome the 
difficulties, Landau and Lifshitz (L-L) first derived a modified R-R equation by replacing $\dddot{\bm{x}}$ in Eq.~\eqref{eq:al} with 
$\frac{\dot{\bm{F}}_{ext}}{m_{e}}$, where the pathological solutions are explicitly avoided in form\cite{Landau1975,Griffiths2010}. But 
we should carefully understand this model, since the intrinsic connection between acceleration and $\bm{F}_{ext}$. In fact, both A-L and 
L-L can be unified into an extended charge model. With a spherically symmetric shell charge distribution, these two classical electron 
models are obtained in limits of infinitesimal charge and slowly varying $\bm{F}_{ext}$ respectively\cite{Levine1977,Blanco1986,
Harte2006,Griffiths2010,Rohrlich1999,Rohrlich2000,Gralla2009,Linz2014}. A basic corollary of extended charge model tell us the anomalies 
only occur while the charge radius is less than the classical electron radius $r_{c}=\tau_{0}c$, where the classical pictures fail.

In summary, there must be something wrong with a cyclotron electron living in classical world. A challenging question then is: 
\emph{How to build a classical electron model in a first-principle framework}? Catati first constructed an improved Lagrangian for Eq.~\eqref{eq:al} 
$L_{\mathrm{C}}=e^{-\frac{t}{\tau_{0}}}\left[\frac{\tau_{0}}{2}\dddot{\bm{x}}^2-\dot{\bm{x}}\cdot\nabla\bm{F}_{ext}\cdot\ddot{\bm{x}}+\frac{1}{2\tau_{0}}\bm{F}_{ext}^2\right]+\bm{s_1}\cdot\left(\dot{\bm{x}}-\dot{\bm{x}}\right)+\bm{s_2}\cdot\left(\ddot{\bm{x}}-\ddot{\bm{x}}\right)$, 
where $\bm{s_1}$ and $\bm{s_2}$ are two multipliers\cite{Carati1998}. Barone and Mendes introduced another kind of auxiliary variable $\bar{\bm{x}}$ 
named as \emph{image}, and constructed a Lagrangian as $L_{\mathrm{BM}}=m_{e}\dot{\bm{x}}\cdot\dot{\bar{\bm{x}}}+\frac{\tau_{0}m_{e}}{2}\left(\dot{\bm{x}}\cdot\ddot{\bm{\bar{x}}}-\ddot{\bm{x}}\cdot\dot{\bar{\bm{x}}}\right)-V\left(\bm{x},\bar{\bm{x}}\right)$\cite{Barone2007}. 
Furthermore, by using three multipliers, Deguchi \emph{et al.} gave two types of Lagrangians for Eq.~\eqref{eq:ald} with a 
source-like term\cite{Deguchi2015}. These Lagrangians imply that a proper R-R action may not be built without introducing 
non-physical degrees of freedom (DoF). Though Eqs.~\eqref{eq:al}-\eqref{eq:ald} can be obtained via stationary variation 
naturally, adjoint auxiliary dynamical equations are also generated, which lead to pseudophysics. Here we suggest a non-local 
Lagrangian without auxiliary variables $L=e^{-\frac{t}{\tau_{0}}}\left[\frac{1}{2}\dot{\bm{x}}^2-\tau_{0}^{-1}\bm{x}\cdot\int^{t}_{0}\bm{F}_{ext}\left(t',\bm{x},\dot{\bm{x}}\right){\rm{d}}t'\right]$, which gives Eq.~\eqref{eq:al} while an electron is immersed in a uniform field. 
Although it has practical value, we emphasize that the Lagrangian we construct is far from complete, since the non-local Lagrangian 
structure is non-physical in classical world. Then another question raises: \emph{Is it applicable to arbitrary field configurations}? 
We have not proved it, and recommend it as an open question which need light to shed on.

\paragraph{Quantum mechanical self-consistent field}---
\label{pa:3}
Due to the simple and intuitive image, the classical electron model is widely accepted by physicists, engineers and 
general populations. But unfortunately, no one has observed single classical cyclotron electron directly. Even more, 
with the help of a novel radio-frequency spectrometer, the indirect detection of cyclotron radiation emissions from a 
mildly relativistic electron has been realized recently\cite{Asner2015}. \emph{Just why do we believe electrons 
are charged point-like objects or small rigid bodies in classical world}? If we can find a good analogue in quantum 
mechanics, we think it maybe a satisfactory answer. After all, our universe is constructed in a quantum form. Let us 
recall a basic fact that the Gaussian wavepacket has a minimum uncertainty product. In other words, it is the maximum 
entropy state on an infinity open interval in all domains. It is an interesting subject which can be understood in different 
profiles. The Gaussian distribution is obviously the only mathematical function that has the same form in a pair of canonical 
conjugate representations. With some potential, it has an invariant shape during evolutions, which is recognized as a coherent 
state. \emph{That is exactly what we are searching for}. Following Schwinger and Glauber's work on optic coherent 
states\cite{Schwinger1953,Schwinger1953-2,Glauber1963,Glauber1963-2}, Malkin \emph{et al.} and Feldman \emph{et al.} first 
constructed the electron coherent states in a uniform magnetic field\cite{Malkin1969,Feldman1970}. 
Given a specified magnetic field $\bm{A}=\left(-\frac{1}{2}By,\frac{1}{2}Bx,0\right)$ in Landau gauge, a magnetic coherent 
state (MCS) can be generated by coherent superposition of the Landau levels $\psi_{n,m}$\cite{Landau1930,Landau1977,Kowalski2005,Zhu2017},
\begin{eqnarray}
\phi_{w_{0},r_{0}}={\rm{e}}^{-\frac{|F|^2+|G|^{2}}{2}}\sum^{\infty}_{n=0}\sum^{\infty}_{m=0}\frac{F^{n}}{\sqrt{n!}}\frac{G^{m}}{\sqrt{m!}}\psi_{n,m},\label{eq:2}
\end{eqnarray}
where $F\equiv{-i\sqrt{\frac{eB}{2\hbar{c}}}w_{0}}$ and $G\equiv\sqrt{\frac{eB}{2\hbar{c}}}r_{0}$ are two auxiliary functions. 
$w_{0}=x_{0}+iy_{0}$ and $r_{0}=r_{x}-ir_{y}$ are wavepacket and guiding centers of the corresponding classical gyro-motion. 
With the help of gauge and coordinate transformations, $r_{0}$ can be dropped and the gyro-radius is $|w_{0}|$. The MCS wavefunction 
can be explicitly given by,
\begin{eqnarray}
\phi=\frac{1}{\sqrt{2\pi\Delta_{z}}\Delta}{\rm{e}}^{\!-\!\frac{\left(r\!-\!w_{0}\right)^{2}+\left(rw^{*}_{0}-c.c.\right)}{4\Delta^{2}}},\label{eq:3}
\end{eqnarray}
where $r=x+iy$. Above function describes a Gaussian wavepacket with width $\Delta=\sqrt{\frac{\hbar{c}}{eB}}$ on the X-Y plane. 
Due to the electron along magnetic field is unbounded, the longitudinal wavepacket width $\Delta_{z}$ gives rise to longitudinal 
charge density. In Schr\"{o}dinger picture, $\phi$ does not diffuse and the wavepacket center moves as $w_{0}(t)=w_{0}e^{-i\omega{t}}$ 
with classical gyro-frequency $\omega=\frac{eB}{m_{e}c}$. A classical gyro-motion picture is found via the observables 
$\left<x\right>=\frac{v}{\omega}\sin\left(\omega{t}+\alpha\right)$, 
$\left<y\right>=\frac{v}{\omega}\cos\left(\omega{t}+\alpha\right)$, 
$\left<\pi_{x}\right>\equiv\left<p_{x}-\frac{e}{c}A_{x}\right>=\frac{eB\left<y\right>}{c}$, 
$\left<\pi_{y}\right>\equiv\left<p_{y}-\frac{e}{c}A_{y}\right>=-\frac{eB\left<x\right>}{c}$, 
where phase $\alpha$ determines the initial wavepacket center $\left(x_{0},y_{0}\right)$. Non-trivial mechanical momentum 
$\left<\bm{\pi}\right>$ indicates there is a distributed current associated with the rotating wavepacket. These $\Delta$-independent 
observables $\left<\bm{x}\right>$, $\left<\bm{p}\right>$ and $\left<\bm{\pi}\right>$ equal to the classical ones exactly.

What a beautiful model the MCS is. Exact, simple and clear. But unfortunately the perfect MCS is just a phantom. From above discussion, 
we know that the rotating wavepacket must emit radiations. As a result, the MCS wavepacket can not keep invariant, since radiation damping. 
On a quantum level, we can treat this problem by using the S-M self-consistent field theory naturally. On the cotangent bundle 
$T^{*}G=(\psi_{R},\psi_{I},\bm{A},\phi,\psi_{I}/2,-\psi_{R}/2,\bm{Y},0)$, the S-M Hamiltonian is given by\cite{Qiang2017},
\begin{subequations}
\begin{eqnarray}
H&=&\frac{1}{2}\int\left[\psi_{R}\hat{H}_{R}\psi_{R}+\psi_{I}\hat{H}_{R}\psi_{I}-\psi_{R}\hat{H}_{I}\psi_{I}+\psi_{I}\hat{H}_{I}\psi_{R}\right.\nonumber\\
&&\left.+4\pi{c}^2\bm{Y}^{2}+\frac{1}{4\pi}\left(\bigtriangledown\times\bm{A}\right)^{2}\right]\mathrm{d}^{3}x,\label{eq:hm1}\\
\hat{H}_{R}&=&\frac{1}{2}\left(-\frac{\hbar}{m_{e}}\bigtriangledown^{2}+\frac{e^2}{m_{e}{\hbar}c^{2}}\bm{A}^{2}\right),\label{eq:hm2}\\
\hat{H}_{I}&=&\frac{e}{2m_{e}c}\bigtriangledown\cdot\bm{A}+\frac{e}{m_{e}c}\bm{A}\cdot\bigtriangledown.\label{eq:hm3}
\end{eqnarray}
\end{subequations}
Then the dynamical equations can be written in a canonical formalism as $\dot{S}=\left\{S,H\right\}$. $S$ indicates arbitrary 
DoF on $T^{*}G$.

In atomic units, $\hbar=m_{e}=e=1$ a.u. and $c\approx137$ a.u. A strong background 
magnetic field $\left(0,0,B\right)$ leads to multi-scale non-linear problems. Let $\frac{c}{B}$ 
be a perturbation parameter $\epsilon$, the asymptotic serieses can be defined as 
$\widetilde{\bm{A}}=\sum\limits^{\infty}_{n=0}\epsilon^{n}\bm{A}_{n}$, 
$\widetilde{\psi}_{R}=\sum\limits^{\infty}_{n=0}\epsilon^{n+\frac{1}{2}}\psi_{Rn}$, 
$\widetilde{\psi}_{I}=\sum\limits^{\infty}_{n=0}\epsilon^{n+\frac{1}{2}}\psi_{In}$. Then 
a class of undimensional asymptotic equations are obtained by substituting these serieses into 
Hamiltonian \eqref{eq:hm1} $\widetilde{H}=\sum\limits^{\infty}_{n=0}\epsilon^{n}H_{n}$. 
With matched asymptotic expansions, the leading order is given by,
\begin{subequations}
\begin{eqnarray}
\epsilon^{0} & : & \ddot{\bm{A}}_{0}+c^2\bigtriangledown\times\bigtriangledown\times\bm{A}_{0}=0,\label{eq:9}\\
\epsilon^{\frac{1}{2}} & : & \dot{\psi}_{0}=\frac{1}{c}\bm{A}_{0}\cdot\bigtriangledown\psi_{0}+\frac{i}{2}\bigtriangledown^{2}\psi_{0}-\frac{i}{2c^{2}}\bm{A}^{2}_{0}\psi_{0}.\label{eq:10}
\end{eqnarray}
\end{subequations}
It is found that the leading order equations describe the background magnetic field and perfect MCS $\psi_{0}=\phi$.

The next-to-leading order is derived straightforwardly,
\begin{subequations}
\begin{eqnarray}
\epsilon^{1} : \ddot{\bm{A}}_{1} & + & c^2\bigtriangledown\times\bigtriangledown\times\bm{A}_{1}=\frac{4\pi}{c}\bm{\mathcal{J}}_{0}.\label{eq:12}\\
\epsilon^{\frac{3}{2}} : \dot{\psi}_{1} & = & \frac{1}{2c}\bigtriangledown\cdot\bm{A}_{1}\psi_{0}+\frac{1}{c}\bm{A}_{0}\cdot\bigtriangledown\psi_{1}+\frac{1}{c}\bm{A}_{1}\cdot\bigtriangledown\psi_{0}\nonumber\\
& & +\frac{i}{2}\bigtriangledown^{2}\psi_{1}-\frac{i}{2c^{2}}\bm{A}^{2}_{0}\psi_{1}-\frac{i}{c^{2}}\bm{A}_{0}\cdot\bm{A}_{1}\psi_{0}.\label{eq:13}
\end{eqnarray}
\end{subequations}
It is found that the next-to-leading order equations describe the radiations induced by perfect MCS current and 
associated radiation corrections for MCS electron. These equations cut off the R-R effects with primary physics. 
$O(\epsilon^{2})$ expansions can be introduced via a same procedure and solved order by order.

Let us examine Eqs.~\eqref{eq:12}-\eqref{eq:13}. The perfect MCS current can be explicitly given by 
$\bm{\mathcal{J}}_{0}=(y+y_{0},-x-x_{0},0)\frac{1}{4\pi\Delta_{z}\epsilon^{2}}e^{-\frac{1}{2\epsilon}|r-w_{0}|^2}$, 
and the associated radiations are evaluated as 
$\bm{A}_{1}\left(\bm{x},t\right)=\frac{1}{c}\int\frac{[\bm{\mathcal{J}}_{0}(\bm{x}{'},t{'})]_{ret}}{|\bm{R}|}\rm{d}^{3}x{'}$, 
where $\bm{R}=\bm{x}-\bm{x}{'}$, and the retarded bracket can be expanded in 
$[\cdot]_{ret}=\sum\limits^{\infty}_{n=0}\frac{(-1)^{n}}{n!}(\frac{|\bm{R}|}{c})^{n}\frac{\partial^{n}}{\partial{t}^{n}}[\cdot]_{t{'}=t}$. 
In $\Delta\to{0}$ limit, it reduces to the Li\'enard-Wiechert potential of a classical point-like electron, and the 
Larmor power can be derived straightforwardly. We rewrite Eq.~\eqref{eq:13} in a compact form $\dot\psi_{1}=\hat{H}_{0}\psi_{1}+\hat{H}_{1}\psi_{0}$, 
where $\hat{H}_{0}=\frac{1}{2}\bm{\pi}^{2}_{0}$ and $\hat{H}_{1}=-\frac{1}{c}\bm{A}_{1}\cdot\bm{\pi}_{0}+\frac{i}{2c}\bigtriangledown\cdot\bm{A}_{1}$. 
Then the quasi-classical electron dynamical equation is obtained,
\begin{eqnarray}
{\rm{D}}_{t}\left<\bm{\pi}\right> &=& -\frac{1}{c}\left<(\bigtriangledown\times\bm{A}_{0})\times\bm{\pi}_{0}\right>+\frac{1}{c}\left<(\bigtriangledown\epsilon\bm{A}_{1})\cdot\bm{\pi}_{0}\right>_{0}\nonumber\\
& & -\frac{1}{c}{\rm{D}}_{t}\left<\epsilon\bm{A}_{1}\right>_{0}-\frac{1}{c}\left<(\bigtriangledown\times\epsilon\bm{A}_{1})\times\bm{\pi}_{0}\right>_{0}+o\left(\epsilon^2\right),\label{eq:17}
\end{eqnarray}
where $<\cdot>_{0}$ means a MCS expectation. Let us examine Eq.~\eqref{eq:17}. The 1st term is a Lorentz force induced 
by background magnetic field, the 2nd and 3rd terms make up the 1st-order electric R-R force, and the 4th term is the 1st-order 
magnetic R-R force. With the instantaneous static assumpation, magnetic contributions for R-R can be dropped. The R-R force is 
finally evaluated by,
\begin{eqnarray}
\bm{F}_{R} &\approx& \sum^{\infty}_{n=0}\frac{(-1)^{n+1}}{n!c^{n+2}}\frac{2}{3}\frac{\partial^{n+1}}{\partial{t}^{n+1}}\left<\bm{\pi}_{0}\right>\gamma_{n},\label{eq:18}
\end{eqnarray}
where $\gamma_{n}=\int\int\psi^{*}_{0}(\bm{x})\psi_{0}(\bm{x})|\bm{R}|^{n-1}\psi^{*}_{0}(\bm{x}')\psi_{0}(\bm{x}')\rm{d}^3x\rm{d}^3x'$.
Eq.~\eqref{eq:18} is a quantum analogue of the extended charge model. In the classical model, with a spherically 
symmetric shell type charge distribution, $\gamma_{n}$ can be evaluated explicitly, and the A-L and L-L equations 
are obtained in different limits\cite{Levine1977,Griffiths2010}. If we recognize an electron in quantum mechanical 
self-consistent field theory is a charged finite wavepacket, the self-energy divergence no longer exists. But we 
emphasis that the runaway and preacceleration would emerge from the quasi-classical dynamics if the Compton wavelength 
was less than the classical electron radius\cite{Moniz1974,Moniz1977}.

From classical to quantum, we find a good approach to describe the dynamics of single cyclotron electron with R-R. On the 
contrary, we show an intrinsic connection between two electrons living in classical and quantum worlds respectively. A unified 
physical picture is built. A significant advantage is that everything in this framework is obtained self-consistently, and 
one can check what's wrong with a classical electron model.

\paragraph{Physics beyond the perturbation}---
\label{pa:4}
Although the results of asymptotic analysis provide us with a good picture to unify perspectives on a cyclotron electron in 
different fields, we want to know more. With the help of state-of-the-art geometric simulator designed for S-M systems\cite{Qiang2017}, 
we have access to abundant non-linear and non-perturbative dynamical features of single cyclotron electron in quantum mechanical 
self-consistent field framework. Numerical experiments are implemented on a 400$\times$400$\times$2 uniform Eulerian lattice, where 
the periodic and absorbing boundaries are introduced to cut-off the electron and radiations respectively. In fact, the longitudinal 
periodic condition indicates there are infinite coherent electrons form a cyclotron electron-wire. To keep a tolerable numerical 
dispersion, the lattice scale $\delta_{s}=0.05\Delta$. The temporal step $\delta_{t}=\frac{0.5\delta_{s}}{\sqrt{3}c}$ is adopted 
to achieve a precise dynamical sampling. 

\begin{figure*}
\includegraphics[width=18cm]{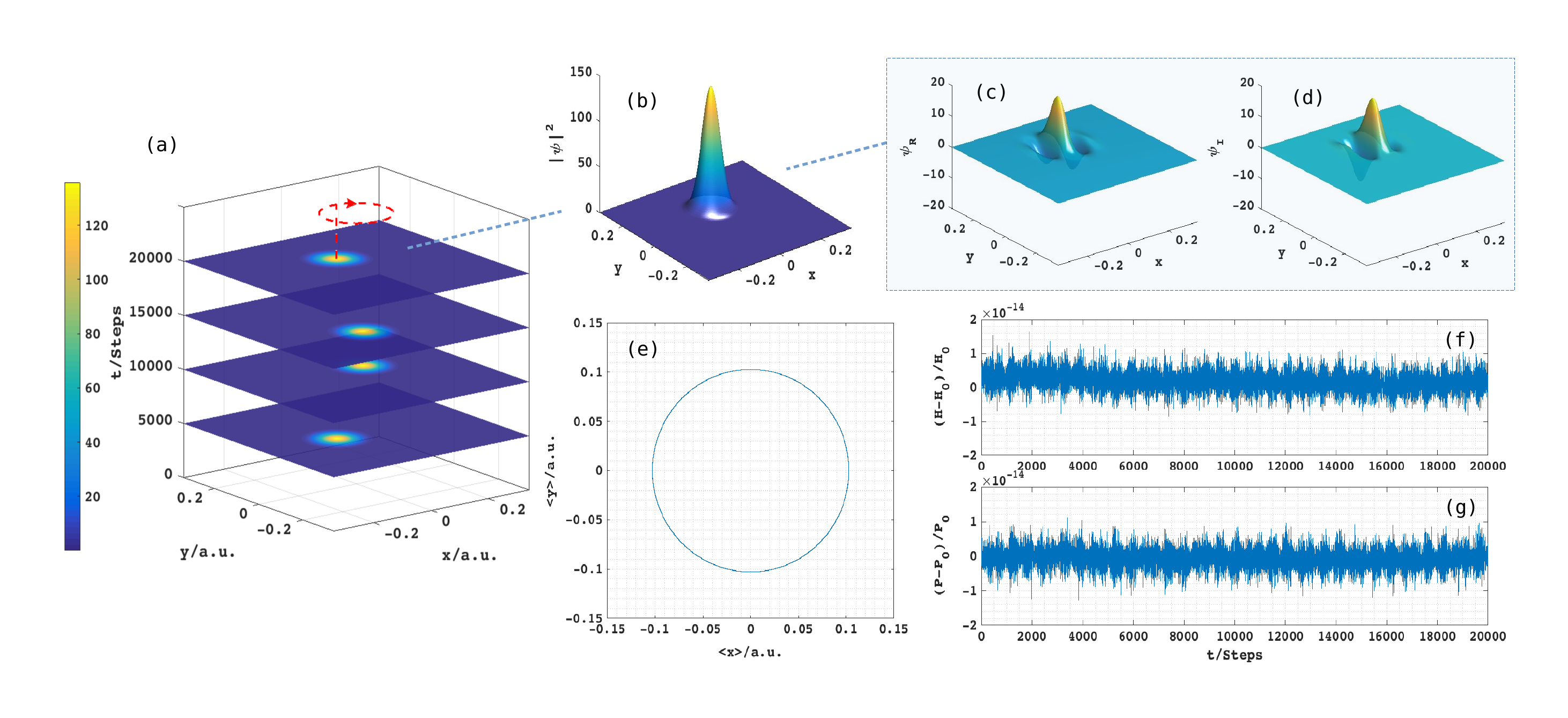} 
\caption{\label{fig:1} A perfect MCS. (a) traces a clear evolution of the MCS. (b) is the MCS wavepacket. 
(c) and (d) show the real and imaginary parts of the MCS amplitude. A non-spreading Gaussian wavepacket 
keeps in the whole life of gyro-motion. A quasi-classical trajectory $<\bm{x}>$ can be found in (e), where the 
orbit keeps a perfect circle. Based on error plots (f) and (g), both total Hamiltonian and total probability errors 
are well bounded. These results provide a benchmark for MCS simulations with R-R.}
\end{figure*}

First, we close the Maxwell's dynamics and observe a perfect MCS in a strong background field $B\approx1.165\times10^{5}$ a.u. 
($2\times10^{12}$ G), where $\Delta\approx3.429\times10^{-2}$ a.u. ($1.814\times10^{-10}$ cm) and $\omega\approx8.504\times10^{2}$ a.u. 
($3.517\times10^{19}$ rad$\cdot$s$^{-1}$). The initial gyro-radius $|w_{0}|=3\Delta$ means the electron velocity $|\bm{v}|\approx87.47$ 
a.u. ($1.914\times10^{10}$ cm$\cdot$s$^{-1}$) and Lorentz factor $\gamma\approx1.299$. After a $2\times10^{4}$ steps (about 10 cycles) simulation, 
the nice results shown in Fig.~\ref{fig:1} print out a dependable and intuitive picture of a non-dissipative cyclotron electron, 
which can be recognized as a good benchmark for following simulations of MCS with R-R.

\begin{figure*}
\includegraphics[width=18cm]{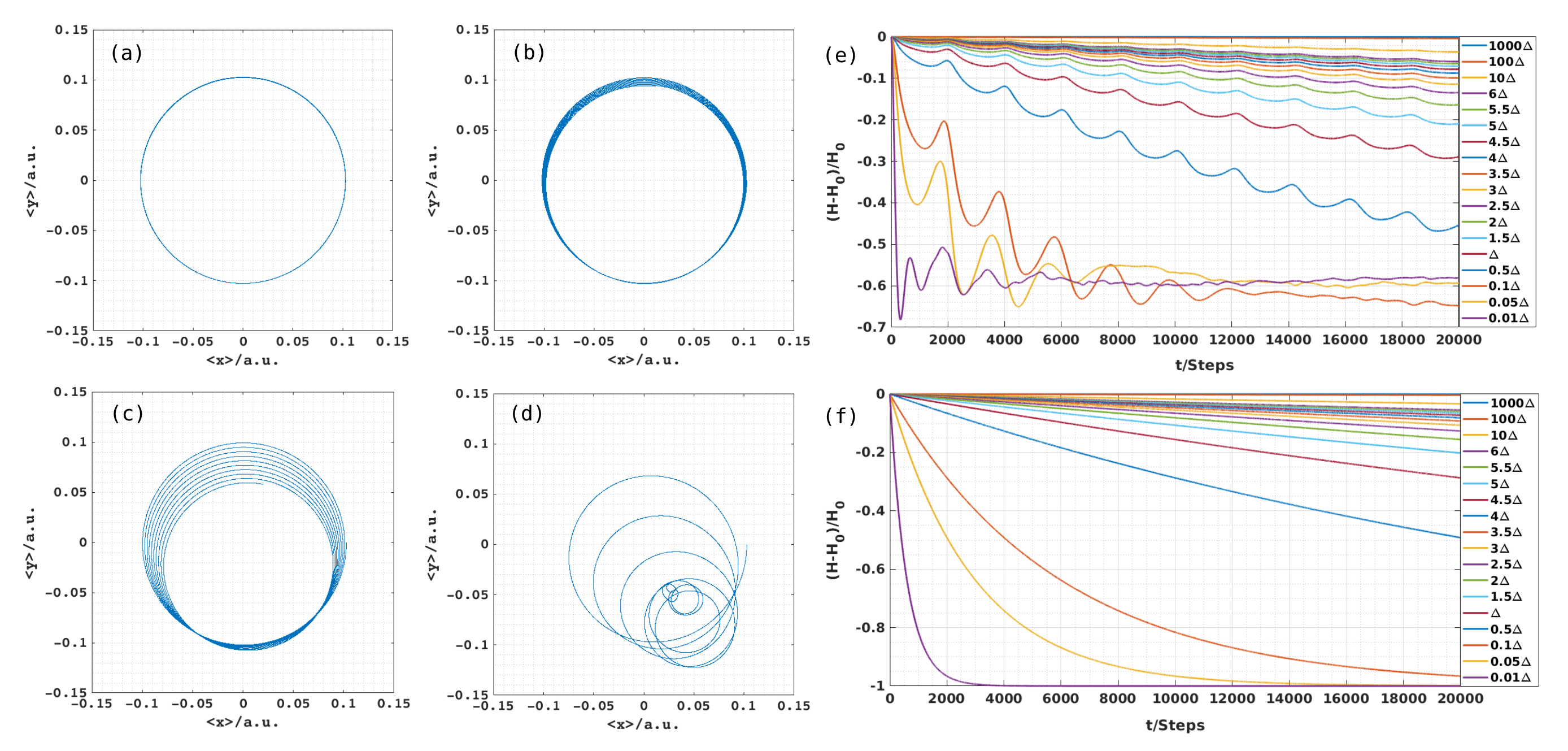} 
\caption{\label{fig:2} R-R induced cyclotron electron decay. (a) - (d) show the quasi-classical trajectories of a MCS 
electron with $\Delta_{z}$ values $5\Delta$, $3\Delta$, $\Delta$ and $0.1\Delta$ respectively. With a large $\Delta_{z}$, 
the orbit decay slowly, where the guiding center is almost invariant and the gryo-radius shrinks uniformly. on the contrary, 
a strong radiation damping leads to observable guiding center drift. In $\Delta_{z}\to{0}$ limit, the trajectory can no 
longer keep a classical orbit and tends to a quasi-random motion around a fixed position. (e) and (f) are electron decay spectra 
obtained by simulations and Eq.~\eqref{eq:20} respectively, which give an optimal complementary relation between classical and 
quantum worlds. The damping curves at weak radiation regions plotted in two insets have a good consistence with each other. The 
discrepancy occurs at weak radiation regions originate from the quantum nature of a cyclotron electron.}
\end{figure*}

\begin{figure}
\includegraphics[width=9cm]{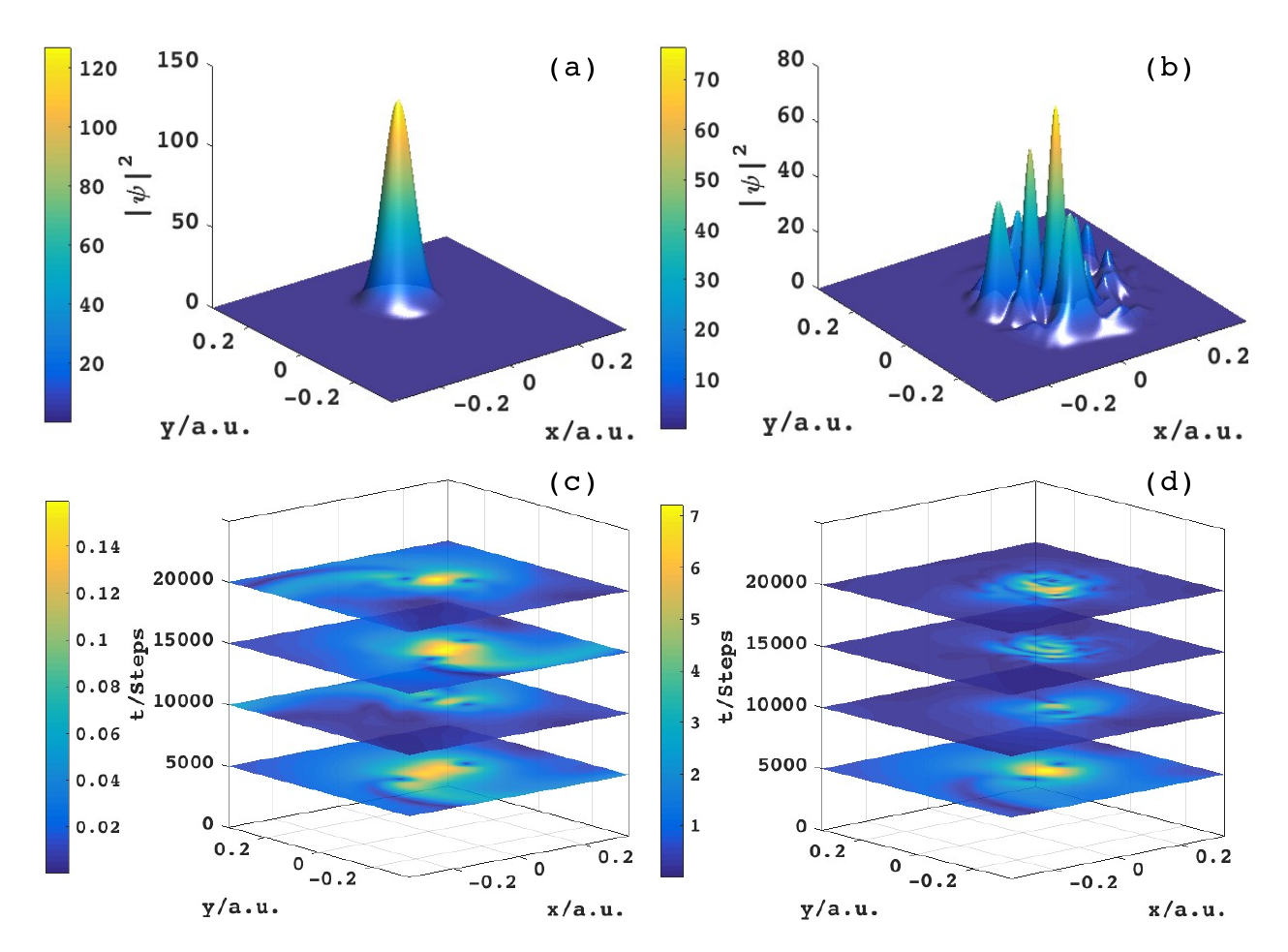} 
\caption{\label{fig:3} MCS VS chimera state. At a weak radiation region ($\Delta_{z}=3\Delta$), 
a MCS keeps a quasi-Gaussian wavepacket, where the final state shown in (a) is a quasi-MCS with 
weak radiative correction. (c) shows the evolution of radiation intensity $\bm{A}^{2}$ ($\Delta_{z}=3\Delta$), 
which illustrates a cyclotron radiation picture. The principal radiation direction along electron 
velocity rotates with the MCS. At a strong radiation region ($\Delta_{z}=0.1\Delta$), a cyclotron 
electron comes into a quasi-steady state which have a coherent envelope and chaotic characteristics. 
The final state found in (b) is a localized random wavepacket which can keep a secular stability. 
From (d) shows the evolution of $\bm{A}^{2}$ during conversion from MCS to chimera state. After $1\times10^{4}$ 
numerical steps, the radiation intensity has lost cyclotron features and a localized adjoint gauge 
field is left.}
\end{figure}

Let us look into the detailed dynamics of a MCS electron with R-R. \emph{How does the features change when R-R is introduced}? 
As shown in Fig.~\ref{fig:2}(a) - (d), the quasi-classical gyro-orbits with $\Delta_{z}$ in different values $5\Delta$, $3\Delta$, 
$\Delta$ and $0.1\Delta$ are plotted orderly. At a weak radiation region ($5\Delta$ \& $3\Delta$), the gyro-orbits decay slowly, 
where the R-R forces can be treated as perturbations, and the classical electron models can be introduced to describe the expected 
dynamical quantities. It is found that the guiding centers are almost invariant and the gryo-radii shrink uniformly, which are typical 
features of a quasi-linear damping oscillator. At an intermediate radiation region ($\Delta$), a distinct feature is the drift motion 
of the guiding center. Just as a non-linear damping oscillator, the averaged damping forces on left and right half-spaces are inequality, 
which lead to the oscillator center drifts. The initial phase induced symmetry breaking is a dominate cause of the drift direction. At a 
strong radiation region ($0.1\Delta$), the electron traces obtaind from $<\bm{x}>$ can not keep classical orbit secularly. After a few 
cycles, the cyclotron electron decays into a class of quasi-steady states, where the traces indicate it tends to a quasi-random motion 
around a fixed position. Now let us recheck Eq.~\eqref{eq:12}, the far field of 1st-order radiation can be evaluated by,
\begin{eqnarray}
\left|\bm{A}_{1}\right|\left(\bm{x}\right)\propto\frac{1}{c\Delta_{z}\sqrt{|\bm{R}|}},\label{eq:19}
\end{eqnarray}
where $|\bm{R}|>>\Delta_{z}$. The limit $\lim\limits_{\Delta_{z}\to\infty}|\bm{A}_{1}|=0$ can be derived from Eq.~\eqref{eq:19} obviously. 
In fact, the near field of 1st-order radiation also obeys this limit because of $\lim\limits_{\Delta_{z}\to\infty}\frac{\ln{\Delta_{z}+1}}{\Delta_{z}}=0$. 
It indicates that when $\Delta_{z}$ approaches infinity, the radiation fields vanish. On the contrary, when $\Delta_{z}$ approaches infinitesimal, 
the radiation fields are divergent. As a quantum extension of a classical cyclotron electron-wire with R-R, $\Delta_{z}^{-1}$ plays the same role of 
linear charge density. Fig.~\ref{fig:2}(e) provide us with a complete electron decay spectrum. As a classical comparison, the mean radiation power 
of a cyclotron electron in a classical electron-wire is approximately evaluated by,
\begin{eqnarray}
P_{ew}=\frac{\sqrt{\pi}\bm{a}^2e^2}{c^2\omega\Delta_{z}},\label{eq:20}
\end{eqnarray}
and the relevant electron decay spectrum is plotted in Fig.~\ref{fig:2}(f). By contrasting these two spectra, we obtain a good complementary 
relation between classical and quantum cyclotron electrons. The spectral lines above 0.5$\Delta$ shown in Fig.~\ref{fig:2}(e) - (f) fit together 
well, which indicate a basic fact that the self-force dressed MCS wavepacket can be accepted as a satisfactory physical picture for classical 
cyclotron electrons at wide weak radiation regions. The spectral lines below 0.1$\Delta$ shown in Fig.~\ref{fig:2}(e) - (f) exhibit very different 
dynamical fectures, which remind us where the frontiers of a classical world emerge. A classical cyclotron electron at strong radiation regions 
will rapidly throw out most of the kinetic energy via radiations. But a MCS wavepacket at same regions seems very stingy with radiations, and a 
considerable rest energy is finally bounded in the following quasi-steady electron chimera state, which indicate there are more interesting fectures 
hidden behind the cyclotron electrons.

Before concluding this letter, we want to detailedly talk about the strange and inexplicable chimera states of a cyclotron electron observed in 
these numerical experiments. Different from quasi-coherent motions of a MCS in the weak limit, non-linear effects at a strong radiation region bring 
many new features which can not be described by classical electron models. As shown in Fig.~\ref{fig:3}(b) ($\Delta_{z}=0.1\Delta$), the final 
state of MCS is a localized random wavepacket which has a coherent envelope and distinct chaotic characteristics. As a control group, the final 
state of MCS ($\Delta_{z}=3\Delta$) is a quasi-Gaussian wavepacket with weak radiative corrections. The chimera state shown in Fig.~\ref{fig:3}(b) 
can keep a secular stability. Although a classical cyclotron electron in a strong magnetic field will stay at a fixed position after a long-term 
radiation, the chimera state can not be recognized as a quantum analogue. By checking the decay spectrum shown in Fig.~\ref{fig:2}(e), we find 
that the chimera wavepacket still keep about $40\%$ initial energy. As shown in Fig.~\ref{fig:3}(d) ($\Delta_{z}=0.1\Delta$), the evolution of 
radiation intensity during conversion from MCS to chimera state illustrates that a localized chaotic gauge field is left together with the 
electron wavepacket when the cyclotron radiations fade away. On the contrary, the evolution of radiation intensity shown in Fig.~\ref{fig:3}(c) 
($\Delta_{z}=3\Delta$) keeps the typical cyclotron radiation features in the whole life of gyro-motion where the principal radiation direction 
along the electron velocity. The transition from MCS to chimera state shows typical weak turbulence features where the wave break occurs following 
the spatial coherent wavepacket continuously\cite{Zhang2021}. Pattern indicates that the bifurcation produces numerous high dimensional topological 
rings and the topology mixing plays a dominate role in the chimera state formation. The increasing short components in phase space leads to more and 
more complex and random waves. After the cascaded ring topologies are broken by topology mixing, the turnulence will be fully formed. These 
electron chimera wavepackets can be recognized as a class of electron-photon quasi-particle states without classical correspondences, which provide 
us with a new perspective on cyclotron electrons on a quantum level.

\paragraph{Outlook}---
\label{pa:5}
To conclude, we have shown a unified physical picture for a cyclotron electron. It allows to 
link different electron models in a strong magnetic field both on classical and quantum levels. Together 
with the advanced geometric numerical tool, the non-perturbative cyclotron dynamics can be theoretically 
studied self-consistently. A detailed investigation of optimal complementary relation between classical 
and quantum cyclotron electrons breaks the wall that separates classical and quantum worlds. Finally, a 
new door towards cyclotron electron relevant researches is opened up.

Furthermore, the strange chimera states observed in numerical experiments at a strong radiation region 
exhibit more inexplicable features hidden in a cyclotron electron. The experimental observations of 
these states can be an interesting open question, which may bring many new directions for future research, 
such as new steady accelerator modes, subatomic storage structures, and quantum logical units. 


\begin{acknowledgments}
This work is supported by the National Nature Science Foundations of China (NSFC-11805273, 11905220, 12005141). 
Numerical simulations were implemented on the SongShan supercomputer at National Supercomputing Center in Zhengzhou, 
and the ShenMa high performance computing cluster at Institute of Plasma Physics, Chinese Academy of Sciences.
\end{acknowledgments}

\providecommand{\noopsort}[1]{}\providecommand{\singleletter}[1]{#1}%

\end{document}